\documentclass[12pt,preprint]{aastex}

\newcommand{\be}{\begin{eqnarray}}
\newcommand{\ee}{\end{eqnarray}}

\begin{document}

%\title{Baryon Anisotropy within Gamma-Ray Burst Jets}
\title{A Model for Fast Rising, Slowly Decaying Subpulses
      in $\gamma$-ray Bursts}
\author{David Eichler and Hadar Manis}
\affil{Department of Physics,
Ben-Gurion University, Beer-Sheva 84105, Israel;
eichler@bgu.ac.il}
%\author{Donald C. Ellison}\affil{Physics
%Department, North Carolina State University, Box 8202, Raleigh, NC
%27695, U.S.A.; don\_ellison@ncsu.edu}

\begin{abstract}
Gamma ray bursts (GRB's) often feature subpulses that have a
distinctively asymmetric profile -- they rise quickly and decay much
more slowly, while their spectrum softens slightly with observer
time. It is suggested  that these subpulses are caused by slow
baryonic clouds embedded within a primary $\gamma$-ray beam, which
scatter the $\gamma$-radiation  into our line of sight  as they
accelerate. Good quantitative agreement is obtained with observed
light curves and spectral evolution. The kinetic energy that the
baryonic component of GRB jets receives from the primary
$\gamma$-radiation is predicted to be about equal to the amount of
$\gamma$-radiation that is scattered, consistent with observations
of afterglow. Several other observational consequences are briefly
discussed. The possibility is raised that the time scale of short
GRB is established by radiative acceleration  and/or baryon
injection rather than the time scale of the central  engine.
\end{abstract}


\section{Introduction}

The nature of the central engine of GRB's remains unresolved by
observations. It is not known  what the primary form of energy
outflow is, or how the $\gamma$-radiation near the spectral peak is
powered. Popular models invoke a baryonic outflow as the primary
form of  energy output, and synchrotron emission from electrons that
are shock-accelerated  in optically thin regions of the outflow
(Meszaros \& Rees 1994), presumably by internal shocks caused by the
unsteadiness of the flow. Other models suggest that the primary
output of the central engine is photons (e.g. Eichler 1994, Eichler
\& Levinson 2000, Rees \& Mezsaros 2005) or Poynting flux (e.g.
Thompson 1994, Lyutikov \& Blandford 2003). If the observed
$\gamma$-radiation (and presumably a residue of pairs) is the
primary form of released energy at the photosphere of the outflow,
then the question is if and how the $\gamma$-radiation accelerates
the baryonic outflow responsible for GRB afterglow.

In this {\it Letter} we propose that many $\gamma$-ray bursts are
viewed at an angle that is slightly offset from the direction of
the primary $\gamma$-ray beam, and that much  of the primary
radiation is, consequently, observed only after being scattered by
 slower  baryons that have not yet reached their terminal
 Lorentz factor $\Gamma_{\infty}$. The existence
of slow baryons in the flow is not a strong assumption.  They could
enter the primary beam from the periphery due to the diffusion of
slow neutrons from a surrounding wind (Eichler and Levinson 1999;
Levinson and Eichler 2003) or possibly by turbulent excitation of
transverse motion by Kelvin--Helmholtz instabilities at the side.
Moreover, radiatively driven instabilities could bunch baryons
within the outflow. Even in the internal shock model, slow material
would be necessary to dissipate the energy of fast material, and the
shocked material following their collision would still be slower
than at the eventual terminal Lorenz factor. The main assumption
made here is that the photons, though they may scatter off clouds of
high optical depth, are not trapped {\it within} a large optical
depth, and that they accelerate the clouds/bunches as they overtake
them from the rear.
 We show that the observed temperature, as
defined by the location of the emission peak, should then decrease
approximately as $t^{-2/3}$, as reported by Ryde (2004). We also
show that the observed time profile has a characteristic fast rise,
slow decay typical of sub-pulses\footnote{often referred to as
'FRED's -- for fast rise, exponential decay} often seen in the
prompt emission of GRB's. The decay is due largely to the
acceleration of the scattering baryons by the radiation pressure of
the primary photons, which causes the beam of scattered photons to
narrow to below the offset angle.

It would not be surprising if matter were accelerated beyond the
photosphere because the isotropic equivalent luminosity can be as
high as $10^{15}$ Eddington luminosities or more.  This would
surely be a powerful accelerator of any material that did not
already have a Lorentz factor of at least several hundred,  even
at distances of $10^{13}$ to $10^{14}$ cm, where the photosphere
is placed in many models. The acceleration of isolated baryons
would be much faster than the hydrodynamical timescale. In the
process of being accelerated, the baryonic scattering material
would scatter as much radiative energy as the energy it received
from the radiation pressure (see below). Thus, radiation scattered
by accelerating  baryons would represent as significant a part of
the GRB energy budget as the baryons.\footnote{Note that even in
the internal shock model, where the exiting photons are secondary
products of fast baryons overtaking and shocking slightly slower
ones, the resultant super-Eddington photon output would
pre-accelerate most of the slower baryons before they were
overtaken. A self-consistent treatment of this model should
therefore include such scattering and pre-acceleration as well.}

%The possibility is also briefly considered that
%the central engines of  short GRB have
%the same lifetime as those of long GRB, and that the difference
%in the observed duration is due to the difference in the proximity
%of the scattering baryons to the source.

The idea that most GRB's are observed from an offset viewing angle
is well motivated and  has been amply discussed in the literature.
One motivation is that very bright GRB's such as 990123 are much
brighter than the detection threshold and could be viewed by
observers that are offset by an angle of  several times $1/\Gamma$,
just from their kinematic broadening. Another is that the Amati [and
Ghirlanda] correlations (Amati et al 2002, Ghirlanda et al 2004) can
be explained as kinematically correlated softening of both the peak
frequency and isotropic equivalent [jet] energy (Eichler and
Levinson 2004, Levinson and Eichler 2005, Eichler and Levinson
2006). Scattered radiation from accelerating baryons can, at some
cost to peak brightness, significantly enhance the solid angle over
which a GRB could be detected, and, for sources with low
$V/V_{max}$, thus enhance the maximum volume $V_{max} \equiv \int
r_{max}^3(\theta,\phi)d\Omega/3$ over which a burst can be detected.
A considerable fraction of all detected GRB's, or subclasses
thereof,  may then involve a component of scattered radiation. Many
observers could detect such radiation while the baryons that
scattered it into the line of sight continue in a slightly different
direction, and this separation between baryon direction and prompt
$\gamma$-ray direction  could be a reason for the delay of strong
afterglow in many GRB's (Eichler and Granot 2006). We will also note
that at viewing angles $\theta_V$ that are large compared to the jet
opening angle $\theta_{\circ}$, $\theta_V \gg \theta_{\circ}$, this
scattering can give rise to outliers from the Amati relation.

Very recently, Pe'er et al (2007) have independently  proposed that
the softening of the spectral peaks of GRB subpulses results from
viewing material that is, with observer time,  increasingly further
in angular separation from the line of sight. This must always occur
if the jet is extended in angle, but gives a steeper decline of
spectral peak frequency with time ($t^{-\alpha}$, $\alpha \sim 1$)
than the model proposed here, as well as a steeper decline in
luminosity.

\section{A Simple Model}

To calculate the luminosity of scattered radiation as a function of
time, we assume that the scattering material is introduced into the
primary photon beam at rest or with purely radial motion, and that
the size of the photon source is negligible. The photons are thus
completely combed radially and their flux in the scatterer frame is
$F^\prime = \frac{1-\beta}{1+\beta}F$ (Landau and Lifschitz 1962),
where $F$ is the flux in the source frame at radius $r$.  Assue
first, for convenience, that the cloud is optically thin. The
acceleration in the scatterer frame is given by
\begin{equation}
d\beta_s/d\tau = F^\prime \sigma_T/m_ic^2 =
\frac{1-\beta}{1+\beta}F\sigma_T/m_ic^2
\end{equation}
and thus
\begin{equation}
d\beta/dt_{source} = \Gamma^{-3}d\beta_s/d\tau = \Gamma^{-3}F^\prime
\sigma_T/m_ic^2 = \Gamma^{-3}\frac{1-\beta}{1+\beta}F\sigma_T/m_i
c^2. \label{dbdt}
\end{equation}
Integrating this equation in the late time limit
 %behavior of $(1-\beta)$,
  when $ (1+\beta)\sim 2$, $Fdt_{source}\propto d\beta/(1-\beta)^{5/2}$,
  we obtain   $(1-\beta) \propto (\int
Fdt_{source})^{-2/3}$. The received flux by a distant observer at
viewing angle $\theta_V$ as a function of observer time $t$ is given
by (Rybicki and Lightman 1979)
\begin{equation}
dP_e/d\Omega = [\gamma(1-\beta \cos\theta_V)]^{-4}dP^\prime
/d\Omega^\prime ,
\end{equation}
and the late time behavior when $1/\Gamma \ll \theta_V$ is, for
constant F,  $dP_e/d\Omega \propto (1-\beta)^3 \propto t^{-2}$.

Note that for a powerful GRB  with isotropic equivalent luminosity
$L_{iso} = 10^{53}L_{53}$ erg/s, and baryons beginning from rest
at radius $r = 10^{12}r_{12}$ cm, the acceleration time up to
$\Gamma \le  (10^{8}L_{53}/r_{12})^{1/3}$ is  less than the
hydrodynamical expansion time (Eichler 2004) for $r_{12}, L_{53}
\sim 1$.   Thus, the model predicts {\it a priori}, given the
relevant range of parameters for the central engine and host star
envelope ($L_{53} \sim 1$, $r_{12} \sim 1$), that the baryons
naturally obtain a Lorentz factor of several hundred.

It is possible that the baryonic cloud is optically thick when
injected into the primary beam. In this case, the back end is
compressed by the radiation pressure and a reverse shock is sent
through the cloud (which could result in particle acceleration and a
nonthermal component in the scattered radiation), and
 the average acceleration is reduced by the optical depth $\tau$.
The scattered radiation emerges from the back end of the cloud,
after only one or very few scatterings, so, in the frame of the
cloud, the forward hemisphere is shaded. {\it When the cloud
accelerates beyond $\Gamma =  1/\theta_V$, the observer's line of
sight emerges from the shadow, and a sudden turn on of the scattered
radiation is seen.} The turn on is just at the value of $\beta$
where the flux of scattered radiation detected by the observer is
near maximum.  Alternatively, the optical depth $\sigma_T
\int^{\infty}_r n(r^{\prime})[1-\beta(r^{\prime})]dr^{\prime}$ of
any given parcel of baryons also drops, due to the acceleration,
much faster than the expansion. The photosphere can  {\it
self-organize} in the sense that a sudden drop in optical depth is
then both the cause and effect of a sudden drop in $[1-\beta]$.
%As
%long as the cloud is optically thick, the scattered radiation must
%then escape backwards in the frame of the cloud, which is
%illuminated from behind, and, in the source frame, all of the
%scattered radiation lies {\it outside} the $1/\Gamma$ beaming cone.
Finally, the cloud may be optically thin.  This would imply that
much of the primary radiation escapes unscattered.  This possibility
seems to be allowed, at present, by observations: A GRB as powerful
as 990123, for example, though only occurring once per $\sim 10^3$
bursts, is  $\sim 10^3$ times as powerful as the  typical GRB, and ,
having a high peak frequency,  is a logical candidate for primary
emission. The proposed model can thus accommodate a rather large
range of initial optical depths, as long as the key assumption is
maintained that the photons are not trapped within the baryons as
they would be if everything
were distributed smoothly. %i.e. before
%they lose most of their energy to adiabatic expansion. This
%renders unnecessary (though does not prohibit) recovery of the
%energy lost to baryons with internal shocks.

\begin{figure}
\epsscale{.90} \plotone{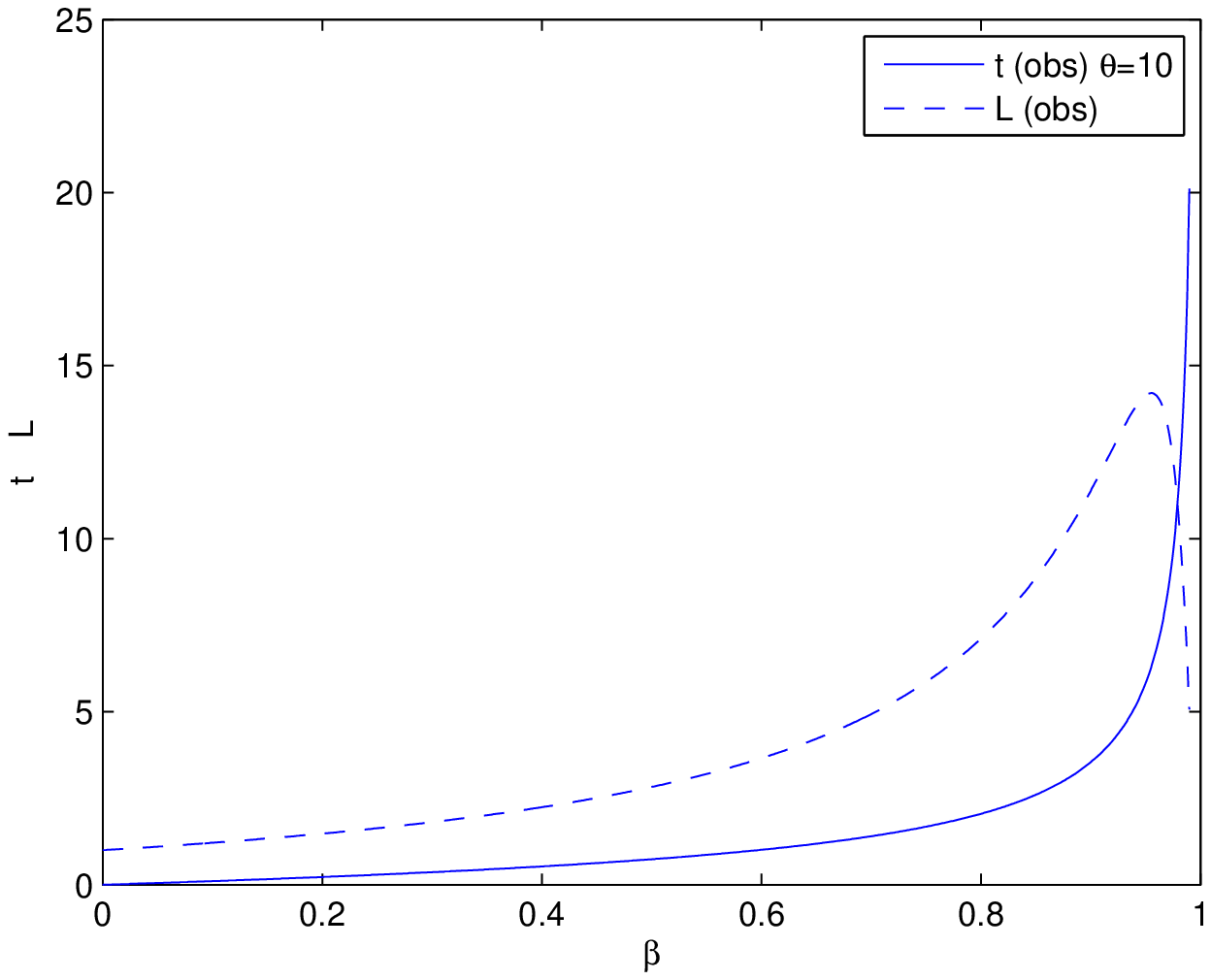} \caption{The luminosity and time,
both as measured by an observer viewing the GRB  at an offset angle
$\theta_V = 10$ degrees, are displayed as functions of $\beta$ of
the scattering plasma. It is assumed that the plasma is optically
thin. The normalization of L and r, the distance of the plasma from
the central engine, are arbitrary. }\label{fig_1}
\end{figure}

\begin{figure}
\epsscale{.90} \plotone{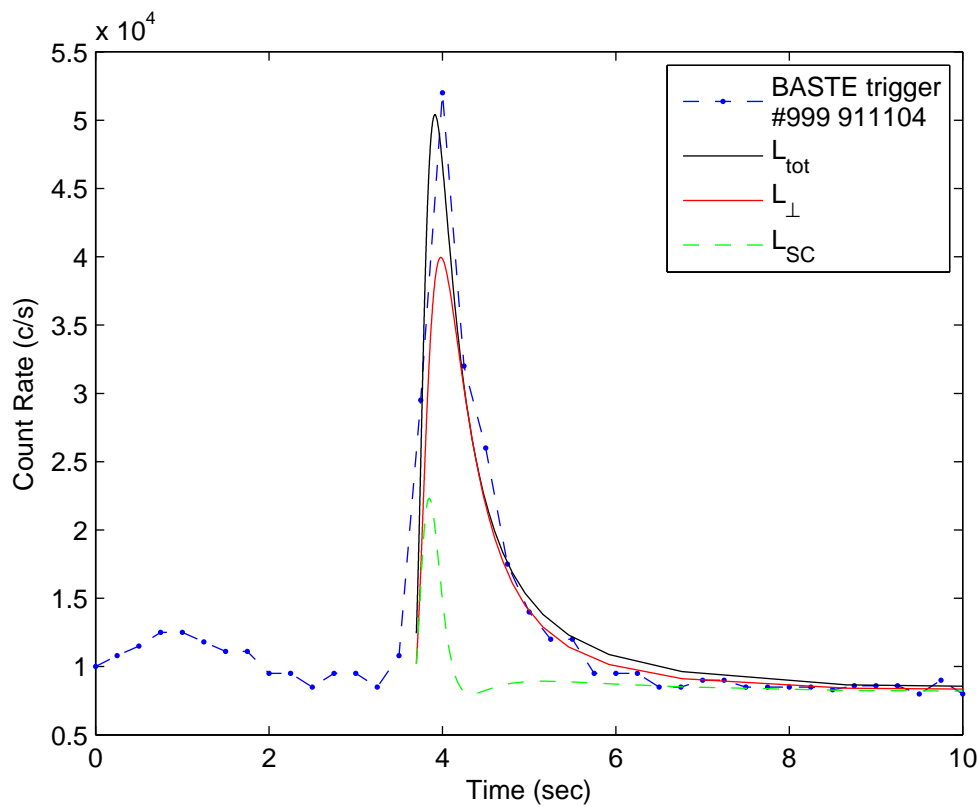} \caption{GRB 911104 fitted by L(t)
as prescribed in Figure 1. The curves labeled $L_{sc}$  and
$L_{\perp}$ show the respective contributions of  the polarizations
in and perpendicular to the scattering plane under the assumption of
an unpolarized primary  pencil beam. The count rate assumes a
spectrum of dN/dE$\propto E^{-1}$, from an assumed  detector
threshold energy $E_{th} = 30 KeV$ to  a maximum energy in the
source frame $E_{max}$ of 2 MeV. \label{fig_2}}
\end{figure}

\begin{figure}
\epsscale{.90}
\plotone{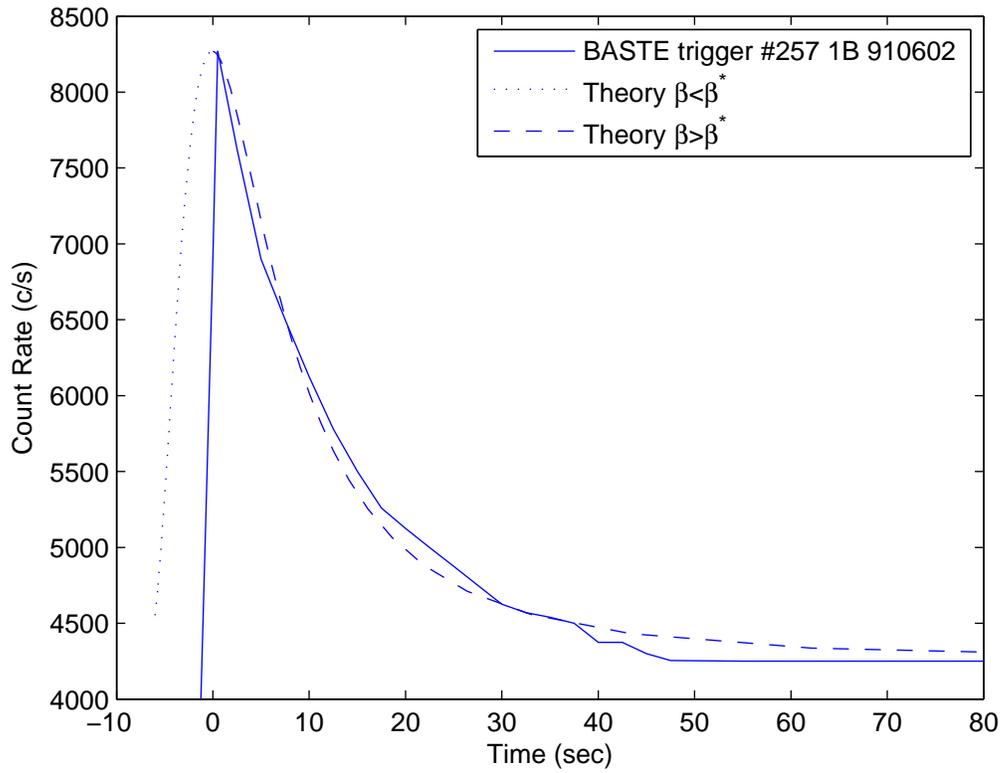}
\caption{GRB 910602 fitted by L(t) as prescribed in Figure 1, but
with the assumption that the scattering material is injected with
  $\beta = \beta^*$. The part of the curve to the left of the observed
rise is therefore not physical.} \label{fig_3}
\end{figure}

 In  Fig.~\ref{fig_1}, we have plotted $\hat t \equiv (\sigma_T/m_ic^2)\int (1-\beta
\cos \theta_V)F dt_{source}$ as a function of $\beta$, the velocity
of the scatterer in units of $c$. Here $F$ is to be taken at the
instantaneous position of the scatterer. We have also assumed a
plane parallel geometry, which is valid when the acceleration is
rapid compared to the expansion time. Assuming constant $F$, we have
then plotted in Fig.~\ref{fig_2} the observed light curve
$dP_e/d\Omega$ as a function of $\hat{t}$. Here, we have assumed
that the baryonic cloud is optically thin, so that its emission can
be seen by the observer even when $\Gamma \ll 1/\theta$. It is seen
even in this case that the characteristic time asymmetry that is the
signature of GRB subpulses is nicely reproduced by the model as seen
in Fig.~\ref{fig_2}.  In the case where the cloud is injected - or
the radiation first transmitted to the observer - at finite $\beta$,
the rise is even sharper relative to the decay, and the peak can be
a true cusp ( Fig.~\ref{fig_3}).

%Because we have selected
%particularly sharp,
%asymmetric subpulses out of a rather varied selection, we have
%tailored our assumptions to give the sharpest, most asymmetric
%peaks.
%We have assumed that the injection of baryons into the
%primary radiation beam is instantaneous - or that the optical
%thinning of the flow is sudden because of very rapid acceleration.
%Neither assumption is perfect, and indeed, the observed rises tend
%to be a bit more gradual than our calculated rises. However, we
%defer detailed quantitative analysis of the sharp rises to late%r
%work. We have also assumed that the polarization of the primary
%radiation is perpendicular to the observer's line of sight;
%unpolarized primary radiation would give a somewhat rounder peak.

The late time scaling behavior of the spectral location of the peak
is easily calculated by noting that the initial peak photon
frequency in the source frame $\nu_{pi}$ is seen in the scatterer
frame as $\nu_{pi}^\prime = \nu_{pi}/\Gamma (1+\beta)$ while the
final peak frequency, as seen by the observer, is $\nu_{pf} =
\nu_{pi}/\Gamma^2(1+\beta)(1-\beta\cos\theta_V)$. This implies that
$\nu_{pf}=\nu_{pi}m_ic^2
\frac{d\beta}{dt}\Gamma/[(1-\beta)F\sigma_T]$ where equation
(\ref{dbdt}) has been used with
 $dt/dt_{source} = (1-\beta\cos\theta_V)$.
%$^2(1+\beta)$and $1-\beta \ll 1-\cos\theta_V =$,
 For F constant,
and  $\beta$ close to unity, the previous result that $(1-\beta)\sim
1/2\Gamma^{2}\propto t^{-2/3} $ implies that
%$\sim2\Gamma^2(1-\beta cos\theta)$, and
\begin{equation}
\nu_{pf} = -\nu_{pi}\frac{dln(1-\beta)}{dlnt}\Gamma/\hat{t}
%\nu_p/2\Gamma^2(1-\cos\theta_v)\propto\Gamma^{-2}\propto (1-\beta)
 \propto t^{-2/3}
\end{equation}
in good agreement with observations (Ryde 2004).

There are several free parameters in the model, including the
viewing angle, the initial velocity  and radius at which the
baryonic cloud  is injected into the jet (or at which it becomes
optically thin), the rate at which baryons are injected into the jet
(e.g. suddenly or gradually) and the inevitable decline of the
primary luminosity with time.  Nevertheless, we believe that the
fast rise, slow decay is a generic feature of both the observations
and theoretical predictions. We find that the peculiar shape  is
insensitive to the viewing angle, the luminosity and radius; these
parameters basically rescale the x and/or y axis. The decline of the
luminosity with time is expected and could be the reason the
theoretical tails are slightly more prolonged than the ones actually
observed (e.g. Fig. 3), but, because the cloud moves almost as fast
as the photons, the Langrangian derivative of the luminosity is
considerably smaller than the Eulerian derivative, so this effect is
likely to be small. By the same token,  there could be a small
component of the primary beam near the viewing angle, which could
make the luminosity decline more gradually with time than if the
primary beam is entirely within a narrow pencil shape.

In the case that the scatterer is optically thin, the model predicts
correlation of polarization with intensity (Fig. 2). If the primary
radiation is unpolarized, and  the cloud is being accelerated
through the Lorentz factor $\gamma= 1/\theta_V$, then the peak of
the subpulse should correspond to $\Gamma \sim 1/\theta_V$. This
corresponds to a $\pi/2$ scattering in the frame of the cloud and
should therefore correspond to maximum polarization.

It is easy to see that any parcel of energy $dE_{\gamma}$ that
emerges as  scattered radiation by relativistic baryons is about
equal to the kinetic energy $dE_b$ that is imparted to baryons as a
result of its scattering: In the instantaneous rest frame of the
baryon, in the limit of elastic Thomson scattering, the scattered
radiation has on the average zero momentum,  $dp_{\gamma}^\prime =
0$, and all its original energy $dE_\gamma^\prime$, while the
scatterer has gotten all the momentum, $ dp_b^{\prime} =
dE_\gamma^\prime/c$, and essentially none of the energy,
$dE_{b}^\prime \sim 0$. In the source frame, the ratio of energy in
scattered photons $dE_\gamma = \Gamma (dE_{\gamma}^\prime + \beta c
dp_{\gamma}^\prime)$ to the energy they impart to the scatterer
$dE_b=\Gamma\beta  cdp_b^\prime$ is $dE_{\gamma}/dE_b = \beta^{-1}
$, so that in the limit $\beta \sim 1$, the energy that ultimately
remains in the scattered radiation is nearly equal to the kinetic
energy that it imparted to the scatterer. {\it This provides a basis
for why the kinetic energy in the baryons, as inferred from
afterglow data, is comparable to that in the prompt emission.}
Allowing for the possibility of only partial coverage, then in fact
a significant fraction of the primary radiation remains unscattered,
in which case  the energy in prompt emission should consistently
exceed the energy inferred from afterglow data. This is consistent
with the results of Eichler \& Jontof-Hutter (2005), which indicate
that the prompt emission, corrected for viewing angle, is typically
3 to 15 times larger than the estimated kinetic energy
(Lloyd-Romming \& Zhang 2004) inferred from the X-ray afterglow
after 10 hours.
% For the case described
%above of an optically thick cloud thinning into a self organized
%photosphere, a more careful treatment of the radiative transfer near
%and beyond the photosphere is clearly desirable, but beyond the
%scope of this letter. More generally, we should assume that baryons
%can be injected at any optical depth (perhaps right at the base of
%the flow) and the solid angle coverage, optical depth of the
%baryons, and their initial velocity at the injection radius may be
%significantly constrained by careful observations and statistics of
%the time behavior of subpulses within GRB. These observational
%constraints should improve with time, and may serve as useful
%diagnostics of the transverse structure of GRB jets and the baryons
%within them.

\section{Implications for GRB Statistical Correlations}

 We can
imagine three classes of observers: a) those in the direct line of
the baryon beam, b) those that are slightly offset from the baryon
beam and c) those that are at large offset angles from the baryon
beam. For simplicity of discussion, we assume here that the baryon
beam is a filled in cone of opening angle $\theta_{\circ}$. The
direct line observers see more or less what is scattered/emitted by
the baryons, appropriately blue shifted by the factor
$\Gamma_{\infty}$. Those slightly offset by angle $\Delta \theta
\equiv \theta_V - \theta_{\circ} \ge 0$, nevertheless have a good
chance to observe the baryons even after they have reached their
terminal Lorentz factor if $\Delta \theta \Gamma_{\infty}$ is not
too large. In contrast to observers in the beam, they see the
emitted radiation blue shifted by only the lower Doppler factor
$D\equiv 1/\Gamma_{\infty}[1-\beta cos(\Delta \theta)]$, and the
total fluence they measure that is contributed by any given pencil
beam scales as $D^3$. In the case of $\Delta \theta \Gamma_{\infty}
\gg 1$, which implies $D\sim 1/\Gamma_{\infty}[1-cos\Delta \theta]$,
and $\Delta \theta \ll\theta_{\circ}$, the fraction of the total
beam that contributes to the radiation in the observer's direction
is proportional to $1-cos\Delta \theta = 1/D$, so that the total
fluence is proportional to $1/D^2$. This is essentially the Amati
relation (Eichler and Levinson 2004). That the dynamic range of the
Amati relation covers two orders of magnitude in frequency suggests
that this interpretation invokes a range of $ \Delta \theta
\Gamma_{\infty} \le 10$.

 Finally, at large $\theta_V$, $\Delta \theta \gg 1/\Gamma_{\infty}$, the
 offset may be too large for the observer to detect any significant
 contribution from  the baryons moving at terminal Lorentz factor.
 However, if the baryons at some point accelerated through the
 Lorentz factors $\Gamma \sim 1/\theta_V \ll \Gamma_{\infty}$, then such an observer could
 nevertheless see a $\gamma$-ray pulse  from the baryons as they were accelerating
 through  the Lorentz factor $\Gamma_V  \equiv 1/\theta_V$ lasting the acceleration time
 at $\Gamma = \Gamma_V$. The peak frequency at the peak of the
light curve is half the intrinsic peak frequency of the source,
independent of $\theta_V$.
 Scattering by slow ($\Gamma \sim 1/\theta_V \ll 1/\theta_{\circ}$)
baryons, in contrast to the viewing angle effect when $\Gamma \Delta
\theta \ge 1 $, widens the observable photon beam relative to the
baryon beam without significantly altering the observed spectrum. It
thus introduces one-sided scatter into the Frail, Amati and
Ghirlanda etc. correlations in that it lowers the observed fluence
(and ultimately the inferred $E_{iso}$ ) %- relative to what an
%observer in the direct path of the primary photons would have
%observed  -
without altering the inferred (via afterglow breaks) opening angle
of the jet or observed  spectral peak. At a given spectral peak,
therefore, there should always be outliers that appear underluminous
in the context of these correlations, or overly hard spectra for a
given $E_{iso}$. This is consistent with observations  (e.g. Butler
et al 2007and references therein ).

The angular profile of the time integrated scattered radiation has,
during the acceleration phase of the scatterer, a "universal"
structure and a systematic correlation between fluence and observed
duration: For any given source, the amount of energy $E(\theta_V)$
that fills a cone of opening angle $\theta_V$ is proportional to
$1/sin\theta_V$ when the observer is well outside the cone (or
annulus) of primary emission $\theta_{\circ}$. This follows from the
fact that the amount of energy scattered by a scatterer at energy
$\Gamma mc^2$ is proportional to $\Gamma$, while the observer at
viewing angle $\theta_V$ does not see any scattered radiation after
the scatterer has accelerated much beyond $\Gamma \sim
1/sin\theta_V$. Because the emission cone's solid angle $ \Omega_V$
at the observed peak of the subpulse goes as $1-cos \theta_V$, the
observer sees a fluence ${\cal F} = E(\theta_V)/ \Omega_V d_l^2$
(where $d_l$ is the luminosity distance) that scales as ${\cal F}
\propto 1/sin\theta_V (1-cos\theta_V)\sim 2/\theta_V^3$, and this
would induce some scatter in  the prompt radiative output as
inferred from the Frail anti-correlation between isotropic
equivalent energy and apparent opening angle (the latter being
inferred from afterglow breaks). Specifically, GRB subpulses
observed at very large $\theta_V$ ($\gg1/\Gamma_{\infty})$  would
appear less energetic, though their spectra might not be especially
atypical.
%between fluence and peak frequency. %Moreover, the decrease in
%apparent radiative output with $\theta_V$ is only for observers with
%large enough $\theta_V$, $\theta_V \gg \theta_{\circ}$, that
%the primary jet can be taken as a pencil beam. %If, on the other
%hand, the jet and embedded baryons are extended in solid angle,
%and the observer is just at the edge, then the scatterer need not
%be too much slower than the primary fireball in order  to scatter
%some of it towards the observer, and  the  scattering does not
%much alter the fluence measured by the observer or the observer's
%estimate of the jet energy.
% Moreover, the Amati and Ghirlanda relations have not, to our knowledge, been
%tested on individual subpulses. The above discussion, therefore,
%does not obviously impact the significance of the Frail, Amati, and
%Ghirlanda relations other than to provide a reason for residual
%scatter and outliers at very large offset viewing angles.}
Low energy GRB such as GRB 980425 might thus lie well off the Frail
relation because they are observed at large viewing angle. In the
case of large viewing angle, the observed duration $\Delta t_{peak}$
of the subpulse peak scales as $\Gamma_{peak} \sim 1/sin\theta_V$
because, although the observed time lapse is compressed as $\Delta
t_{peak}/\Delta t_{peak,source} \sim 1/\Gamma_{peak}^2$, the
acceleration time in the source frame goes as
$[-dln(1-\beta)/dt_{source}]^{-1} \propto \Gamma_{peak}^3$. The
number of photons scattered during this interval, in the limit of
constant source luminosity,  scales as $\Delta t_{peak}$.
 %Finite photon energy and
%flux thresholds, and declining source luminosity all weaken this
%otherwise linear correlation. %This is in good agreement with figure
%2 (upper right) in Butler et al (2007), which shows that
%$n_{\gamma}$ scales as $T_{90}^{\alpha}$, for $\alpha$ slightly less
%than unity.
The peak flux $E_{\gamma}(\theta_V)/\Delta t_{peak}\Omega_V d_l^2$
is proportional to $1/\Omega_V d_l^2 = 1/2\pi(1-cos\theta_V)
d_l^{2}$. Assuming a flux detection threshold independent of $\Delta
t$ and that  $d_{l,max}^2$ scales roughly as $1/ \Omega_V$ (which
neglects non-Euclidian cosmological effects)  the maximum volume of
detectability $V_{max}\sim d_{l,max}^3 \Omega_V$ scales  as $
1/(1-cos\theta_V)^{1/2}$, %Thus, bursts observed at large angle due
%to scattering by baryons at an early stage of acceleration have,
%through that stage, an output of
%scattered radiation%\footnote{Such an
%observer would be unlikely to detect the jet break and by "inferred"
%we mean that such an observer correctly guesses the solid angle
%subtended by the pulse he observes. The guess for a very nearby weak
%burst would likely be that $ sin\theta_V \sim 1$ and that the jet
%energy is therefore close to the isotropic equivalent energy.}
that is only $\sim[(1-cos\theta_{\circ})/(1-cos\theta_V)]^{1/2} $
times that of a typical GRB. This suggests that GRB seen at large
$\theta_V$ due to scattering by slow baryons  should be relatively
infrequent, and not terribly contaminate or obscure general trends
in GRB statistics. However, a
% If there were no other
%issues that bear on the relative frequency offrom this viewing
%angle, would be detected only $
%[(1-cos\theta_{\circ})/(1-cos\theta_V)]^{1/2} $ times as frequently.
%They would tend to be of short duration, underluminous, and hard. A
more detailed analysis of the effects of scattering by slow baryons
on GRB statistics, well beyond the scope of this paper, should
include  empirical estimates of  the relative occurrence of short
duration, hard, underluminous GRB,  and the  fraction of GRB output
that can or needs to be attributed to slow baryons as opposed to
those already at terminal Lorentz factor. The likelihood that GRBs
can have different classes of host stars, and thus different modes
of baryon injection into the fireball can also be a complex matter
that affects the statistics of GRB parameters.
%
%For example, a
%GRB such as GRB 980425, which has an estimated output of several
%times $10^{48}$ ergs (i.e. only $10^{-3}$ of the usual estimate for
%a typical GRB) would be observed only once per $10^3$ bursts, which
%is consistent with recorded observations.
%\propto E_{jet}$, where $E_{jet}$ is the inferred jet energy

\section{Summary and Further Discussion}

We  have  proposed that at least part of the baryons in GRB are
frequently bunched, either because they are injected in bunches, or
because radiatively driven instabilities bunch them. The primary
radiation that scatters off them is seen by an offset observer as
fast rise, slow decay subpulses, of the sort typically seen in
GRB's. The peak frequency as seen by the observer decays roughly as
 $t^{-2/3}$, in
agreement with the data analysis of Ryde (2004). Because this
softening is kinematic, the power spectrum of rapid variations
originating in the source, if they survive time of flight dispersion
due to the finite size of the scattering region, should, even in the
limit of zero scatterer size,  be softened in the same way as the
spectral peak.  This could provide a future confirmation  of the
model. Another possible observational consequence of the model
(cleanest if the scatterers are optically thin)  is that the
polarization should correlate positively with the received flux in
the subpulse.

In constructing the simplest mathematical model for scattering by
primary $\gamma$-rays by slow material, we have assumed that the
scatterer is point-like in solid angle, and that the radiation is
radially combed. In reality, any given observer may see a
superposition of radiation from a finite range of directions,
including primary emission beamed directly at him. Moreover, the
finite angular spread $\Delta \theta$ of the primary photons, which
may not be negligible if and when $\Gamma$ approaches $1/\Delta
\theta$, is determined by the collimation profile imposed by the
host star. These considerations undoubtedly vary from burst to
burst, and can, along with other variable factors, provide the rich
diversity of individual light curves and spectra found among GRB's.

We might even conjecture that the difference between short and long
GRB's lies not so much in the lifetime or energy of the central
engine, but rather in the acceleration time of baryons in the path
of the fireball and/or the duration of baryon injection into the
primary beam. Differences in the apparent acceleration time would
likely accompany the differences in the corresponding host stars
(e.g. a white dwarf or neutron star merger that collapsed into a
black hole would accelerate baryons much closer to the central
engine), and could also be due at least in part to differences in
the observer's viewing angle. This interpretation of short GRB's
would be consistent with the view that they are typically observed
from a much larger viewing angle off the jet axis  than long GRB's
and {\it appear} to have lower energy than long GRB's. It must be
kept in mind, however, that this model for short GRB's would be
constrained by any millisecond variability observed. It would also
be worth looking for "breakout flashes" at large viewing angle,
which could occur when a GRB fireball is just breaking through the
uppermost layer of its post main sequence  host star. The baryons in
the way of the GRB fireball just as it is breaking out would be
accelerated on a timescale much shorter than the hydrodynamical
timescale, so the observed duration would be short. The short term
(millisecond) variability, however, would be washed out by the
scattering, and this would distinguish it from other short GRB's.
Standard short GRB's, by the same token,  may simply be breakout
flashes from merged white dwarfs or neutron stars.

We are grateful to Dr. A. Celloti for calling our attention to the
data analysis of Ryde (2004) and for several  discussions in that
early stage. DE acknowledges the hospitality of the Kavli Institute
at the University of California, Santa Barbara, and organizers of
its program on Astrophysical Outflows at which these discussions
took place. We also thank Dr. Y. Lyubarsky for helpful discussions.
We are grateful to Drs. S. Patel and U. Griv for technical
assistance in downloading and plotting the data.  We acknowledge
support from the US-Israel Binational Science Foundation and the
Israeli Science Foundation's Center of Excellence Program.

{}

\end{document}